\documentclass[runningheads]{llncs}
\usepackage{graphicx}
\begin{document}
\title{A Geo-Gender Study of Indexed Computer Science Research Publications}
%
%
\author{
	Belén Vela\inst{1}  \and
    José María Cavero\inst{1}  \and
    Genoveva Vargas-Solar\inst{2}  \and
    Javier A. Espinosa-Oviedo\inst{3} \and
    Paloma Cáceres\inst{1} 
}

\authorrunning{B. Vela et al.}
%

\institute{
    Universidad Rey Juan Carlos\\%
    \email{\{paloma.caceres, belen.vela, josemaria.cavero\}@urjc.es}
\and
    CNRS, LIRIS-LAFMIA\\ 
    \email{genoveva.vargas-solar@liris.cnrs.fr}
\and
    Univ. Lyon 2, ERIC-LAFMIA\\
    \email{javier.espinosa-oviedo@univ-lyon2.fr}
}

%
\maketitle              
\begin{abstract}
    This paper presents a study that analyzes and gives quantitative means for measuring the gender gap in computing research publications. The data set built for this study is a geo-gender tagged authorship database named authorships that integrates data from computing journals indexed in the Journal Citation Reports (JCR) and the Microsoft Academic Graph (MAG). We propose a gender gap index to analyze female and male authors' participation gap in JCR publications in Computer Science. Tagging publications with this index, we can classify papers according to the degree of participation of both women and men in different domains. Given that working contexts vary for female scientists depending on the country, our study groups analytics results according to the country of authors affiliation institutions. The paper details the method used to obtain, clean and validate the data, and then it states the hypothesis adopted for defining our index and classifications. Our study results have led to enlightening conclusions concerning various aspects of female authorship's geographical distribution in computing JCR publications.

    \keywords{
        Geo-gender Study \and 
        Research Publications \and 
        Computer Science \and 
        MAG dataset \and
    }
\end{abstract}

\section{Introduction} 
\label{sec:intro}

Women comprise a minority of the Science, Technology, Engineering and Mathematics (STEM) workforce, particularly in computer science, physics, and math. Quantifying the gender gap may help identify fields that will not reach parity without intervention, reveal under-appreciated biases, and inform benchmarks for gender balance among conference speakers, editors, and hiring committees. 

Productivity and impact measures are the established and reference indexes to evaluate the quality of scientists’ careers. Despite the controversy, descriptive statistics-based indexes are used to evaluate scientific careers. For the sake of objectivity, these measures are adjusted with pondering factors that “consider” some contextual aspects modelling the conditions in which cold numbers are computed (e.g. discipline, hierarchical position, career duration, etc.). 
However, it seems that parenthood, particularly motherhood, marriage and divorce, gender balance in research groups, disease, economy, countries political situation, are factors with a high impact on scientists productivity. 
Many groups and organizations discuss these questions and try to give quantitative numbers about these aspects that finally determine the period in which scientists continue to be competitive and how do social factors influence the standard deviation \cite{LaVanguardia2018}. The objective of the analysis proposed in this paper is to conduct a first statistical study focusing on the publications gender gap. 
The idea is to observe whether the gender gap is visible in high impact papers and whether it changes according to the institution location. For instance, is the gender gap proportionally smaller in European paper authorships\footnote{An authorship represents the participation of an author in a paper; thus, one paper has many authorship, one per author.}? Are there any differences between men and women productivity according to the geographical area? Are women or men more likely to publish in certain computing areas than in others? We relate this quantitative observation with contextual data like the notoriety of the journals of the publications. 
We also compare our results with studies that address computer scientists’ gender gap.

from different open datasets, including Microsoft Academic Graph \cite{Microsoft2015}, ORCID \cite{ORCID2018}, the 2016 JCR release with 497 journals \cite{ORCID2018}, the Genderize.io tool \cite{Genderize.io2018} and Google Geolocation API \cite{Google}. Of course, results are estimations and their veracity pondered by probability. We group authorships mainly by country and then by gender, under the hypothesis that both features directly determine scientists' productivity. Other measures regarding personal history, physical age, the institution and group's notoriety, gender gap in professional environments, administrative duties, and parenthood have not yet been considered because of the little information available about personal data.  

The remainder of the paper is organized as follows. Section \ref{sec:related} discusses related work.  Section \ref{sec:analysis} describes the initial data sets and the cleaning and preparation process performed. Particularly it describes the approach used for determining the country of authors’ affiliations and the gender of papers authors. The resulting set is a geo-gender tagged publication data set called authorships. Section \ref{sec:gendergapindex} introduces the gender gap index that we propose and uses it to analyze the authorships dataset,  discussing the results regarding productivity, country, and gender. Finally, Section \ref{sec:conclusion} concludes the paper and discusses future work.

\section{Related work}
\label{sec:related}

Our proposal's related work is three-fold: first approaches used for "discovering" people gender given their name, particularly paper authors; second, approaches "discovering" the country of an institution given its official name; third, gender gap studies. 

\subsection{Discovering gender given authors names}

\cite{LIS-Bibliometrics2010} recommends best references and tools to find author names, APIs and software to analyze gender and discusses how gender can be analyzed using gender bibliometrics searchers. It identifies where to find author names lists, application programming interfaces (API’s) that can search for the gender of an author like Gender API, Namsor, GenderChecker, Genderize.io, the Python package Sex Machine. Two references on finding out a person gender given her name in the context of studies regarding gender equity in science are \cite{Lariviere2013,Elsevier2017}. \cite{Elsevier2017} discusses how to parse Web of Science names data for various countries. The Elsevier report is combined with Scopus Author profiles that include gender-name data extracted from social media, applied onomastics, and Wikipedia. Furthermore, the paper highlights several challenges to consider: assumption of binary gender, and dearth of valuable author metadata and regional name-gender data (i.e., many names- gender analysis tools are biased towards Western names, making it difficult to do accurate analysis on authors from other areas of the world).

Developing an automatic pipeline that considers all factors that determine the associated gender of the first name is a challenging problem. Since our study's objective was not to focus totally on this issue, we decided to use the country and first name even in composed names as discriminant parameters to determine the gender of a person with such a given name.

\subsection{Determining research institutions location country}

Automatically identifying affiliations referring to any particular country for paper authors introduces several significant linguistic challenges. For example, ambiguity in cities names, where a city's name can correspond to different geographical locations and the names themselves, can have different variants. \cite{Torvik2015} proposes an algorithm that addresses these problems. Key components of the algorithm include a set of 24,000 extracted city, state, and country names (and their variants plus geocodes) for candidate look-up and a set of 1.1 million extracted word n-grams, each pointing to a unique country (or a US state) for disambiguation.

Simple but sometimes effective solutions use geolocation services like google location and google maps and other alternative ones. These services provide API's that can be used in a program for sending requests batches and obtain a ranked list of locations. The critical issue is to have clean institutions names and consider that some institutions have campuses in different locations. For paper authors affiliations, this is important if a study aims to identify even the institution of the institution in which an author works. When results contained ranked location lists, it is essential to consider a reasonable hypothesis for choosing the country because the top-ranked answer is not always the "correct" answer.

\subsection{Measuring gender gap in science}
Several works have investigated the gender gap in scientific publications in other disciplines than computer science. 
\cite{Sax2002,Xie1998} analyze the gap in research productivity between female and male scientists working in universities in the United States of America (USA). They also study this gap in US scientists in four prominent nationally representative cross-sectional surveys of postsecondary faculties in 1969, 1973, 1988, and 1993. In \cite{Mauleon2006}, analyze both the research productivity and impact and publication habits of materials science scientists concerning their gender. \cite{Aksnes2011} start from the received idea that “female scientists tend to publish fewer publications than do their male colleagues” and analyze whether gender differences can be found in terms of citations that support or contradict this. Gender gap analysis can also be found in disciplines closer to computer science, like information systems. \cite{Gallivan2006} studied the research productivity of top researchers in information systems journals. They compared their data with previous studies and with an estimated population of information system researchers and observed that 17\% of the top 251 researchers were women.

In \cite{Holman2018} estimate the gender of 36 million authors from more than 100 countries publishing in more than 6000 journals, covering most STEM disciplines over the last 15 years. The study is done using the PubMed and arXiv databases. They propose a Web application allowing easy access to the data \cite{Holman2019}. The study concludes that the gap is enormous in authorship positions associated with seniority, and prestigious journals have fewer women authors. Additionally, it estimates that men are invited by journals to submit papers at approximately double the women's rate. Finally, the study shows that wealthy countries, notably Japan, Germany, and Switzerland, have fewer women authors than those with evolving economies.

In computer science, few gender gap studies focus on the publications, and they address statistics regarding college, master and PhD inscription, graduates and professional evolution. For example, the National Center for Education Statistics \cite{NationalCenterforEducationStatistics2018} has studied the number of female PhD graduates in computer science in the USA from 1969 to 2010 and observed that it was 17.22 \%. Moreover, there are fewer women than men in computer science in higher education \cite{Papastergiou2008} except some like \cite{Gharibyan2006,Othman2006,Ceci2011}, few studies analyze the causes of women underrepresentation in science. The rare studies conclude that discrimination is not the cause of this in some fields of science. Women participation in computer science has been compared to a shrinking pipeline \cite{Camp1997}, in which the ratio of women decreases as regards the number of female students compared to the number of women who hold positions in academia.

In previous work, \cite{Cavero2015} conducted a study on computer science publications from 1936 to 2010 to analyze women's evolution in computing research. The analysis addressed the women participation as authors of publications, productivity and the relationship with the average research life of women in comparison to men, the gender distribution of conference and journal authorships depending on different computer science topics, and authors' behaviour concerning collaboration with one gender and/or the other. 

Women Matter is a series of reports willing to demonstrate the economic benefits encouraged by increasing feminine professional participation. Reports have proposed nine global studies and five regional studies, including Europe, Asia, Latin America, Africa and the Middle East, eight countries USA, Canada, Turkey, UK, India, and Spain. Results show that inclusion and raise economic productivity insist on each region's challenges and strategies to achieve gender diversity in leadership.

The gender gap is a multi-perspective phenomenon that should be analyzed beyond trends analysis. First, even if the trend shows increasing participation of women in science activities, the evolution pace must be discussed. There is no point to be happy that we need 100 years or more to close or significantly reduce the gender gap in science. Indexes measuring and correlating women and men productivity and the way their productivity is correlated with other social and political issues, like the reputation of the institution of affiliation, the country, the domain, scientific public policies, etc. can give more insight into the way gender gap works as a complex social phenomenon. Our work proposed the gender gap index to provide other perspectives of the gender gap in publications with a high reputation in the computing science community.

\section{Data collections analysis}
\label{sec:analysis}

\begin{figure}[t]
        \includegraphics[scale=0.74]{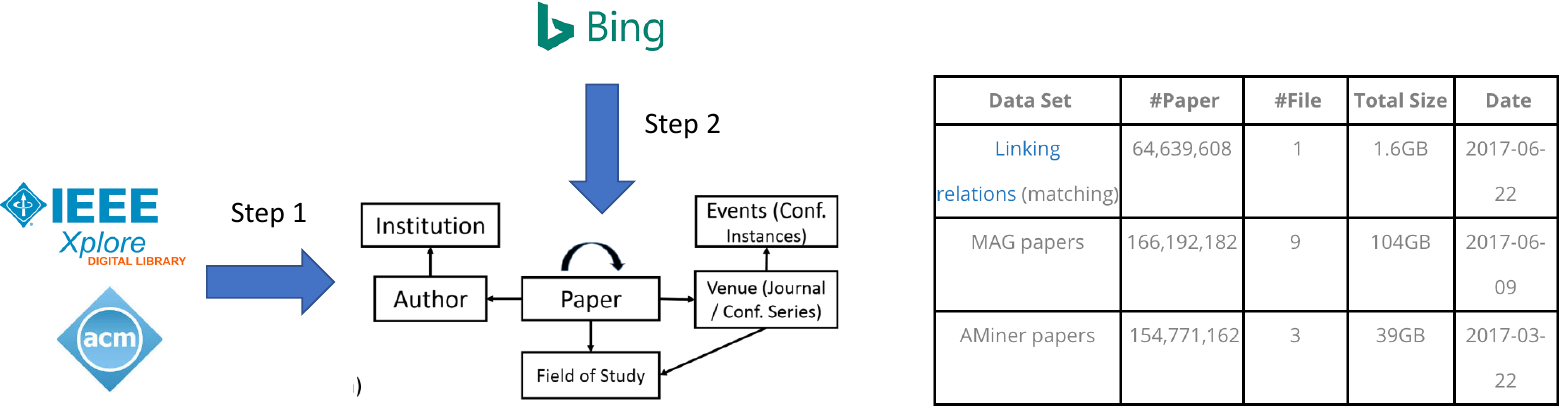}
        \caption{Microsoft Academic Graph structure and content}
        \label{fig:1}
    \end{figure}

According to our research questions, we prepared a dataset from different data collections’ releases providing information about journals, authors publication timeline and authors professional information (i.e., institution, positions and selected publications). We have selected the Microsoft Academic Graph (MAG)\cite{Microsoft2015} which is a heterogeneous graph containing scientific publication records, citation relationships among those publications, and authors, institutions, journals, conferences, and fields of study. 
As shown in figure \ref{fig:1}, the MAG is fed with the IEEE Xplore digital library and ACM publications and completed with content stemming from Bing. It then provides both the graph and synthetic views of its content, including the provider data set, the number of provided papers exported in one or several files, the size of the GB dataset, and the exportation date. For example, the MAG exported the 06/09/2017, 166,192,182 papers in 9 files that represent 104GB.

\begin{figure}[t]
        \begin{center}
        \includegraphics[scale=1.1]{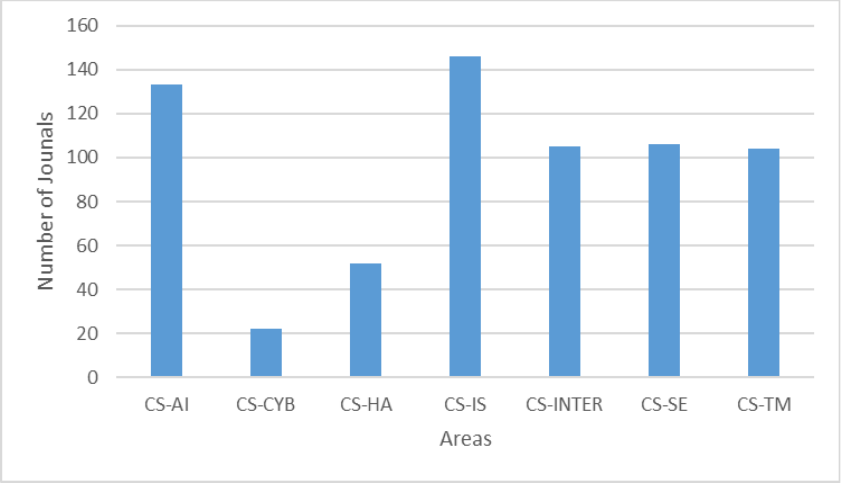}
        \caption{Distribution of JCR journals in Computer Science according to the area}
        \label{fig:2}
        \end{center}
    \end{figure}

Our study focuses on Computer Science journals indexed by the 2016 release of JCR. Therefore, we created a MongoDB\footnote{https://www.mongodb.com} database containing the journals of this JCR release. The result is a set of 668 journals, including the journal's name, the number of cites, the IF number, the Eigenfactor, the year and the area (a journal can be indexed in more than one area). The chart in figure \ref{fig:2}, shows the distribution of journals according to the area, where Information Systems (CS\_IS) and Artificial Intelligence (CS\_AI) are the areas with the most significant number of journals (resp. 146, 133); then Software Engineering (CS\_SE), Interdisciplinary Applications (CS\_INTER) and Theory \& Methods (CS\_TM) with resp. 106, 104, 105 journals; and finally Hardware \& Architecture (CS\_HA) and Cybernetics (CS\_CYB) with resp. 52 and 22 journals. 

\begin{figure}[t]
        \begin{center}
        \includegraphics[scale=1.3]{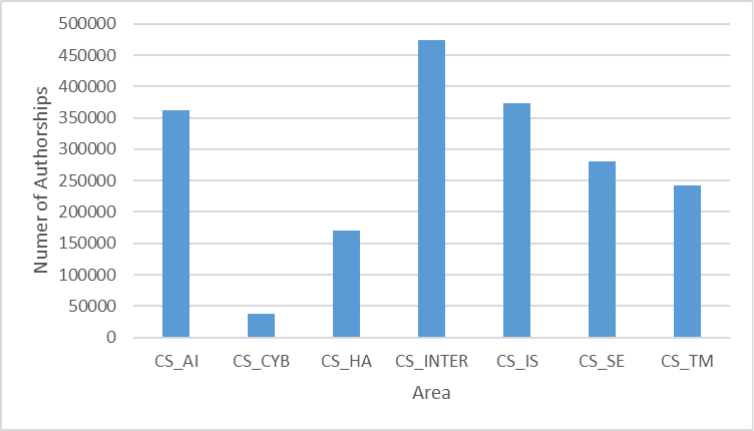}
        \includegraphics[scale=1.3]{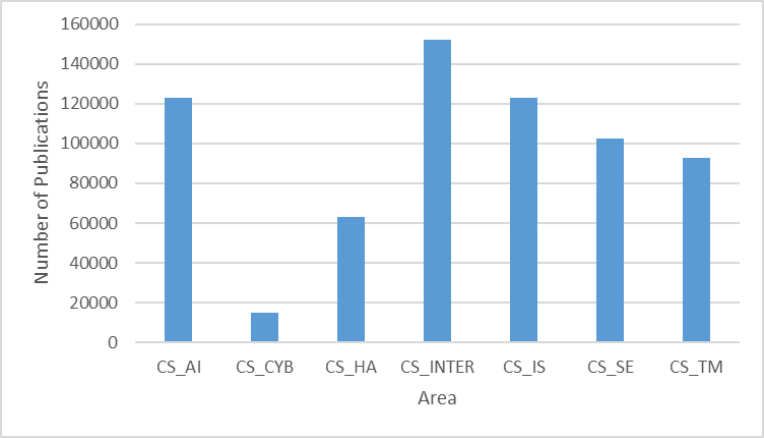}
        \caption{Distribution of publications (a) and authorships (b) per JCR area}
        \label{fig:3}
        \end{center}
    \end{figure}

Guided by the list of journals obtained from JCR, we used the MAG for building a clean Computer Science authorships dataset (see figure \ref{fig:2}). This result includes 1,491,933 authorships of 388 papers published in JCR indexed journals (some appear in more than one JCR area). Each entry in the corpus provides for a given publication, its authors with their affiliations, and the main knowledge domain (i.e., this information comes from the journal). The charts in figure \ref{fig:3} show the statistics of the content.

Our study's viability on the gender gap and our results' quality relied on the authors' names and affiliation quality. Therefore, we performed a data cleaning process on the authorships dataset to reduce noise that could bias the processes of determining institutions country and authors' gender (e.g., special characters, missing values, initials).

\subsection{Determining institutions geographical location}
\label{sub:det-loc}

 Regarding the gender gap in publications, we hypothesise that it varies from the country in which scientists pursue their research. Thus, we had to determine for every paper and every author the country of her affiliation. Using the authorships dataset, we filtered 657,550 different institutions. 
After cleaning the institution's names, we used google maps and google location API to determine its geographical location, especially the country of every affiliation institution. 
Our procedure led to the estimation of the country of 454,054 different institutions and the assignment of the corresponding continent: Asia (AS), Africa (AF), Europe (EU), North America (NA), South America (SA), and Oceania (OC).
Using this information, we combined it with the authorships to compute the productivity in terms of authors' publications and classify them according to the country in which they work.

%
%

\begin{figure}[t]
        \begin{center}
        \includegraphics[scale=1.4]{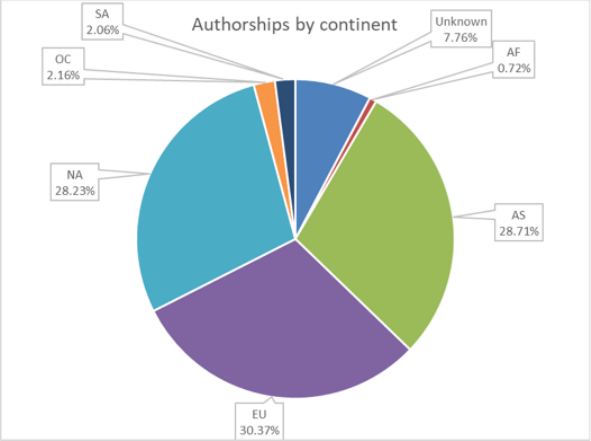}
        \caption{Authorships distribution per continent}
        \label{fig:5}
        \end{center}
    \end{figure}

Figure \ref{fig:5} shows a distribution of the percentage of authorships per continent. The percentage of authorship of European institutions is 30.37\%. The authorships of Asian and North American institutions represent the 28.71\% and 28.23\% of the authorships' dataset. For 7.76\% of the authorships, we could not estimate the institution's country because of special characters' use in the name, for example, "Université de Grenoble III". With another cleaning procedure (automatic or manual), we could have solved the problem. We plan to use classification techniques to classify the noisy substrings in the institution names and then develop a cleaning strategy for each category. However, this concerns future work. 

\subsection{Determining authors gender}
\label{sub:det-gender}

Given that no data collection we had access to gives information about the authors' gender, we estimated it by applying a data analytics method. Using the dataset with 1,491,933 authorships, we projected authors' first names to build a database containing 92,624 different first names and 189,631 different first names associated with the country of their affiliation. Note that we had to exclude 131,955 with only initials as the first name and that for composed names like "José María", for the time being, we just used the first string, thus "José" in our example.

We used the genderize.io tool \cite{Genderize.io2018} to determine authors gender based on their first names. Genderize.io estimates with a certain probability the gender of a given name, including optionally the country's information. This strategy is helpful, as the gender associated with a particular first name may vary according to the country (authors called Andrea are probably female, except if their nationality is Italian, in which case they are probably male). Using the country is tricky for authorships because scientists' nationality is not necessarily the same as their institution nationality (e.g. Mexican scientists working in a French institution). Besides, names are determined by origins or cultural traditions (e.g., Judeo Christian names Josep, José, Joseph, Yohan, generally male names despite the country and the language). A thorough study of these issues will be devoted to future work.

In a first iteration, we used both parameters to estimate the gender of a name: the author's first name and the ID of the country of the institution. We considered valid those estimations with precision between 100-95\%. With this iteration, we estimated the gender of about 31\% of the first names associated with the affiliation institution's country. 

\begin{figure}[t]
        \begin{center}
        \includegraphics[scale=1.3]{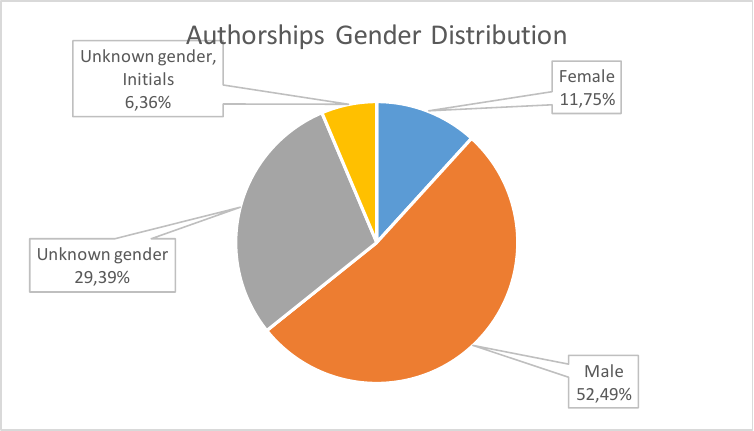}
        \caption{Authorships gender distribution}
        \label{fig:7}
        \end{center}
    \end{figure}

For first names for which we have not obtained a reliable result when estimating the gender with the country parameter, we used only the name to estimate it with genderize.io in a second iteration. With this strategy, we estimated the gender of 53,710 additional names. We then tagged authorships with the gender estimations with precision between 95-100\%. Thus, we tagged with gender about 64,24\% of authorships corresponding to more than 500,000 different scientists. The gender of about 35.76\% (1,443,113) authorships remained unknown (impossible to determine). Of that, 131,955 authorships (6.36\% of the total) correspond to scientists whose name is given in initials. For the remaining authorships’, the genderize.io tool did not provide a trustworthy result (precision less than 95\%). Figure \ref{fig:7} shows the distribution of gender in the authors' dataset.

\subsection{Geo-gender tagged authorships dataset}

\begin{figure}[t]
        \includegraphics[scale=1.3]{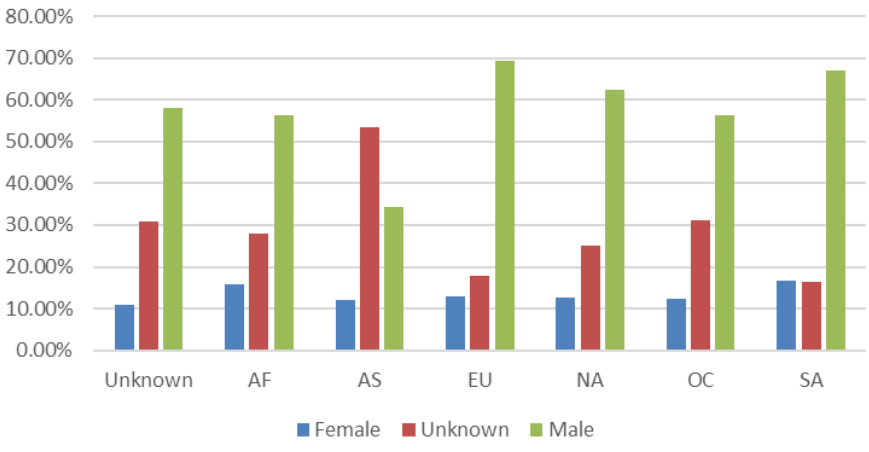}
        \caption{Authorship gender distribution by continent (\%)}
        \label{fig:8}
    \end{figure}

Our study has considered more than 0.5 million publications of the JCR indexed journals and about 1.5 million authorships. Looking for authors gender and countries of institutions applying the methods described in sections \ref{sub:det-loc} and \ref{sub:det-gender}, we ended up with a geo-gender tagged authorships dataset. As seen in figure \ref{fig:8}, the percentage of authorships of unknown gender is higher than 50\% in authors of Asian institutions. In Europe and South America, the percentage of authorships with unknown gender is shorter than 20\%. A sociological study can be applied to understand the cultural reasons that make authors do not write their complete names so that their authorship can be sure for digital engines measuring bibliometry variables. Also, the unknown group can be studied to determine the gender distribution eventually.
To answer this question, we can correlate authorships with ORCID information. Indeed, processing the last name and biography can eventually estimate the authors' gender. Other crowdsourcing methods can also be applied, directly contacting authors that can claim authorship and reveal their gender. This task belongs to our future work.

\section{Gender Gap Index}
\label{sec:gendergapindex}

We used the obtained geo-gender tagged authorships dataset for computing the gender gap index that we propose for answering some of our research questions. 
%
%
%
%
%
%
%
According to this index, we can measure female authors' ratio in a paper concerning male authors' ratio and possible authors with unknown gender. For the first version of the index, we considered that all authors of a given paper have the same degree of participation in its authorship (i.e., we did not make the difference between first and other authors or on the authors' order). Thus, the gender gap index is given by:

    \begin{equation}
        \frac
            {\#female\_authors - \#male\_authors + \#unknown\_authors }
            {\#total\_authors}
    \end{equation}

We computed the gender gap index for the JCR papers in computer science that we initially selected and cleaned in the geo-gender tagged authorships collection. The result illustrated in figure \ref{fig:8} shows the classification of the 518,264 papers according to (1) those with 100\% female/male authors, (2) those with the majority of female/male authors (F+, M+, 51-99\% of female/male authors in the figure), (3) a balanced number of female/male authors (F+M>50\%), (4) unknown because it is not possible to estimate the gap since there are too much unknown authors (>50\%).

    \begin{figure}
        \includegraphics[scale=1.3]{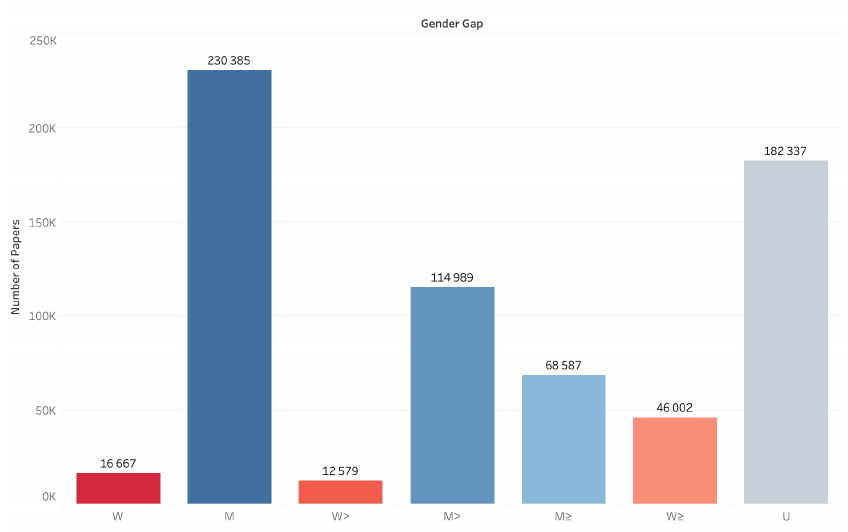}
        \caption{Authorships in computer science research classified according to the gender gap index}
        \label{fig:10}
    \end{figure}

As seen in the previous bar diagram, even if for a good portion of papers, the gender gap is not measurable (140K papers), it is possible to see that the gender gap is significant from several points of view—for example, papers where authors are only male outrage 4,5 times, papers with only female authors. The number of papers with only male authors is ca. 8,5 times higher than papers with totally female authors. There are more papers with the majority of male authors and those with the majority of female authors. The gender gap index can be more useful if combined with the knowledge topic of a paper. 

Regarding the previous information, another interesting question is if women or men are more likely to publish in certain computing areas than in others. Figure \ref{fig:10} shows the distribution of the gender gap classified by the 7 Computer Science areas of  JCR journals: Information Systems (CS\_IS), Artificial Intelligence (CS\_AI), Software Engineering (CS\_SE), Interdisciplinary Applications (CS\_INTER), Theory \& Methods (CS\_TM), Hardware \& Architecture (CS\_HA) and Cybernetics (CS\_CYB). Although no significant differences can be appreciated, Cybernetics and Software Engineering are the two areas with higher participation of female authorships and Hardware \& Architecture is the area with the lowest percentage of women.

\begin{figure}[t]
        \begin{center}
        \includegraphics[scale=1.3]{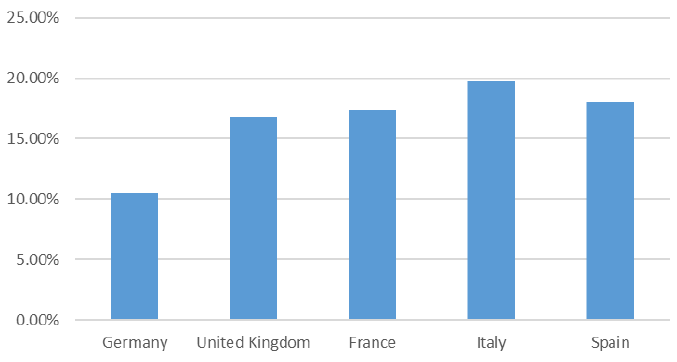}
        \caption{Gender analysis of more productive countries in Europe}
        \label{fig:12}
        \end{center}
    \end{figure}

To figure out if the institution and people's geographic location are correlated, the idea is to observe whether the gender gap is visible in high impact papers and whether it changes according to the institution location. For performing an in-depth analysis of female productivity, we focus in 5 European countries: the United Kingdom, Germany, France, Italy and Spain. Figure \ref{fig:12} shows the percentage of female authorships during the last 50 years. The percentage is calculated as follows:

    \begin{equation}
        \frac
            {\#female\_authorships}
            {\#male\_authorships + \#male\_authorships}
    \end{equation}

We assume that the distribution of unknown authorships is the same as that of known ones. As can be seen, the percentage of female authorships in Italy almost doubles Germany, even if male authorships still outperform female ones in all these countries (see figure \ref{fig:8}).

    \begin{figure}[t]
        \begin{center}
        \includegraphics[scale=1.3]{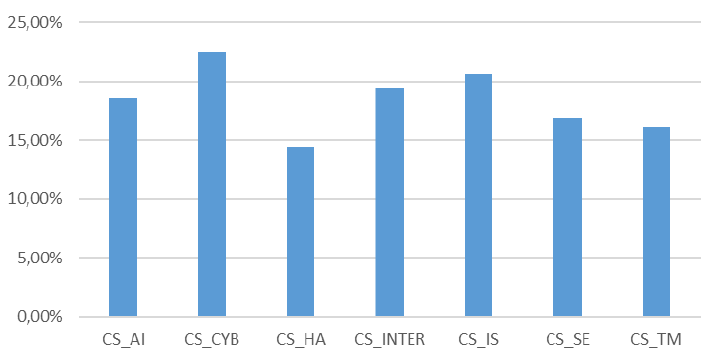}
        \caption{Gender gap distribution according to JCR areas}
        \label{fig:11}
        \end{center}
    \end{figure}

%
%

\section{Conclusion and future work}
\label{sec:conclusion}

This paper proposed the gender gap index to study and measure gender gap patterns in JCR publications in Computer Science. Our main input raw data collections were the Microsoft Academic Graph and a JCR release filtered by domain, i.e., Computer Science. The first result is an authorships data set compiling JCR papers with authors tagged with gender and country of the affiliation institution. The dataset contains a good proportion of non-tagged authorships and lets us develop a first gender gap analysis. We performed a name projection and cleaning pipeline with a strong hypothesis, for example, not considering complete composed names and excluding institutions for which google tools could not directly determine locations. Then, we applied descriptive analytics techniques for providing a first analysis of the publication's practice according to gender and country. This descriptive analysis shows that it is possible to apply the index on those clean gender-tagged authorships (our first result). The gender gap does not show surprising results regarding gender balance; indeed, JCR papers' female authorship shows a significant gender gap in different perspectives. As stated before, the most common and not surprising pattern is that there are more 100 \% male authorships than 100 \% female ones. Close observations show the fact that few papers authorships are total/majoritarian female. 


The gender gap phenomenon in science is complex and calls for research in many different directions. 
For example, it is necessary to propose more effective pipelines for automatically determining authors' gender combining information about their profile in professional networks like Linkedin, Wikipaedia, ORCID, Publons. It is essential to provide publications' gender gap metrics to papers, journals, and editorial boards. If such metadata were included in the published papers, it could be easier to measure the gender gap. Our future work aims at addressing this type of challenges.

\bibliographystyle{splncs04}
\bibliography{GenderGap.bib}

\end{document}